# African Democracy in the Era of Generative Disinformation: Challenges and Countermeasures against AI-Generated Propaganda

Chinasa T. Okolo, Ph.D.
cokolo@brookings.edu
Center for Technology Innovation, The Brookings Institution

## Introduction

As the interest in developing and utilising generative AI grows, there is considerable potential in leveraging these technologies to enhance educational curricula, improve the delivery of healthcare services, and streamline business processes. However, concerns about the negative implications of these technologies have become more prevalent in public discourse, particularly in the context of electoral processes. In light of this prominent discourse, an emerging area of research is investigating the current and estimated impacts of AI-generated propaganda on African citizens participating in elections (Okolo, 2023). Throughout Africa, there have already been suspected cases of AI-generated propaganda influencing electoral outcomes or precipitating coups in countries like Nigeria (Awojulegbe, 2022), Burkina Faso (Smith Galer, 2023), and Gabon (Breland, 2019), underscoring the need for comprehensive research in this domain.

In 2024, over a dozen African countries are scheduled to have elections. Given the existing rate of disinformation that plagues African elections, coupled with social unrest (Salehyan & Linebarger, 2015; Goldsmith, 2015), internet blackouts (Mpofu, 2022), and ballot tampering (Friesen, 2019; Mapuva, 2013), the increasing ubiquity of generative AI heightens concerns that AI-fueled disinformation campaigns could exacerbate current issues within the African continent (Timcke & Hlomani, 2024). With this in mind, concerted efforts are needed to combat such disinformation. Encouragingly, the past decade has led to independent fact-checking organisations throughout Africa. These organisations, like Dubawa in Nigeria, PesaCheck in Kenya, and Africa Check, which operates across South Africa, Kenya, Nigeria, and Senegal, have worked diligently to fact-check widely circulated memes, doctored social media content, and claims made by political figures.

While social media platforms have enacted various safeguards to hinder the proliferation of disinformation, many of these efforts are undermined by loopholes such as the lack of moderation for audio content (Bickert, 2020), as evidenced by recent elections in Slovakia (Meaker, 2023), or automated content moderation systems that have limited understanding of low-resource languages (Nicholas & Bhatia, 2023). Furthermore, citizens with limited digital and media literacy often inadvertently disseminate propaganda (Rubin, 2019). As efforts increase to curb the spread of AI-generated disinformation, governments must allocate resources toward increasing citizen engagement in AI literacy efforts. Governments will also have to establish regulatory frameworks that can effectively govern the use of generative AI and other emerging technologies while promoting safeguards to prevent the spread of disinformation.

This paper aims to highlight the risks associated with the spread of generative AI-driven disinformation within Africa while concurrently examining the roles of government, civil society, academia, and the general public in the responsible development, practical use, and robust governance of AI. To understand how African governments might effectively counteract the impact of AI-generated propaganda, this paper presents case studies illustrating the current usage of generative AI for election-related propaganda in Africa. Subsequently, this paper discusses efforts by fact-checking organisations to mitigate the negative impacts of disinformation, explores the potential for new initiatives to actively engage citizens in literacy efforts to combat disinformation spread, and advocates for increased governmental regulatory measures. Overall, this research seeks to increase comprehension of the potential ramifications of AI-generated propaganda on democratic processes within Africa and propose actionable strategies for stakeholders to address these multifaceted challenges.

## Background literature

Powered by advances in technical research and computing architecture, artificial intelligence (AI) has emerged as a way for people and organisations to create content, optimise workflows, and refine business processes. However, these newfound abilities have increased the harms of AI, introducing new nuances regarding AI-generated content and fueling disinformation. Given the rising concerns about the ability of AI-generated content to impact democratic processes, it is crucial to understand how to actively prevent further harm. To provide sufficient context on the impact of AI on democracy, this section provides a brief overview of generative AI, details the landscape of AI-fueled disinformation throughout Africa, and outlines how AI has shaped democratic processes within African countries.

### Generative AI

Generative AI is a subfield of AI employing advanced machine learning (ML) techniques such as generative adversarial networks (GANs), variational autoencoders (VAEs), and transformers to analyse vast amounts of data to produce content such as text, images, audio, and videos. A prominent form of generative AI is large language models (LLMs), which are AI models trained to interpret linguistic patterns and relationships. Many generative AI methods rely on a range of inputs to generate content, including text, image, audio, and video prompts. Prominent examples of generative AI include text-to-text applications such as ChatGPT, developed by OpenAI; Gemini, developed by Google; and Claude, developed by Anthropic. Popular text-to-image models include DALL-E, by OpenAI, and other tools such as Midjourney and Stable Diffusion. Text-to-video has also seen significant advances with Google's VideoPoet, an LLM that generates zero-shot, variable-length videos and leverages image-to-video functionality and video-to-audio functionally to produce audio output matching video scenes without text guidance (Google Research, 2024). In April 2024, OpenAI released Sora, an AI model that creates hyperrealistic video scenes from text prompts (OpenAI, 2024). In May 2024, Google DeepMind released Veo, a video generation model that leverages prior Google advancements in image and video generation.

While generative AI has been presented as a powerful method to revolutionise how people communicate and create, it is fraught with a plethora of ethical concerns arising from its development, implementation, and usage (Bommasani et al., 2021; UK DSIT, 2024). Vast amounts of data used to train generative AI models were scraped from the internet, mostly without consent. This data includes millions of websites, blogs, research articles, social media posts, books, code repositories, images, videos, audio recordings, and databases. As companies developing AI increase their data needs, the Global South has routinely become a destination for outsourced data labelling labour, with companies such as Appen, Scale AI (Remotasks), Samasource, Clickworker, Mighty AI, Surge AI, Lionbridge AI, and Amazon Mechanical Turk, which played a significant role in the development of notable ML benchmark datasets such as ImageNet, relying heavily on workers in this region. Recent coverage has also highlighted the harms data workers in East Africa have faced when repeatedly exposed to graphic content in their work labelling data for OpenAI's ChatGPT (Perrigo, 2023).

Rising concerns about generative AI have also exposed issues regarding the climate impacts of training large AI models and using generative AI tools, with estimates showing excessive water usage (Crawford, 2024), significant greenhouse gas emissions (Luccioni et al., 2024), and extreme levels of energy consumption (Luccioni et al., 2023). Using generative AI tools has also had many implications within the disinformation landscape. Text and image generation tools have been shown to generate incorrect information (Jo, 2023), spread false narratives (Ferrara, 2024), and manipulate public opinion (Funk et al., 2023; Kidd & Birhane, 2023). Many of the premier generative AI systems are also trained on data that primarily represents dominant languages like English and Western cultures (Shankar et al., 2017; De Vries et al., 2019; Nwatu et al., 2023). Given that content produced by generative AI systems is often hard to detect, such content can easily deceive users who may be unaware of the presence and ability of these tools (Sadasivan et al., 2023; Lu et al., 2023). Thus, it will be essential for researchers, policymakers, and developers to make concerted efforts towards enacting sufficient measures to combat AI-fueled disinformation.

## Disinformation in Africa

Disinformation is the intentional creation and spread of deliberately deceptive information to mislead or influence people, communities, and institutions (Fallis, 2015). African countries particularly struggle with disinformation, with research showing that media consumers in Kenya, Nigeria, and South Africa are exposed to disinformation regularly (Wasserman & Madrid-Morales, 2019). Disinformation has also been an issue long before the internet played a prominent role. It is estimated that from 2000 to 2005, disinformation around the role of HIV in causing AIDS and antiretroviral drugs led to more than 300,000 deaths in South Africa (Chigwedere et al., 2008). Traditional media outlets, including private and state-sponsored platforms, can spread disinformation through radio and television long before it is amplified online (Africa Center for Strategic Studies, 2021). Issues with disinformation have been

Okolo | 3

fueled by greater access to information through social media and other digital platforms (Ceylan et al., 2023). These issues have been further exacerbated by the introduction of generative AI, which has made it easier to create various forms of content (Xu et al., 2023; Chu-Ke & Dong, 2024).

Within Africa, disinformation affects democratic processes along with routine aspects of daily life, such as health and well-being. Research has shown that 20% of media coverage of genetically modified foods by African outlets contains misinformation (Lynas et al., 2022). The COVID-19 pandemic had a major impact on the spread of health disinformation. Repeatedly forwarded messages purporting homoeopathic remedies to prevent and treat COVID-19 were commonly spread through messaging and social media platforms (Nwankwo et al., 2020; Walcott, 2020). Disinformation often appears in cycles, and older news pieces or reports can be brought back into resurgence. In 2020, an article published in 2013 claiming that 50 children in Chad were paralysed by a vaccine supported by the Gates Foundation was debunked (Africa Check, 2020). In 2013, Africa Check was also vital in debunking claims made by the World Health Organization in 1983, stating that 80% of rural Africans primarily depend on traditional medicine for their healthcare needs (Africa Check, 2013). Within Africa, disinformation is often present and rises during elections and periods of social unrest. The recently published book, "Digital Disinformation in Africa" details government-fueled disinformation during the #EndSARS protests in Nigeria, efforts by citizens to counter state disinformation in Angola, disinformation by the Kenyan news media during the 2022 elections, disinformation during the 2018 presidential election period in Cameroon, and how nefarious actors use selfies and hashtags to fuel disinformation during armed conflict in Ethiopia (Roberts & Karekwaivanane, 2024).

While disinformation in Africa is spread by local actors, a growing number of reports have uncovered the rising influence of foreign disinformation campaigns within African countries (Mare et al., 2019; Maweu, 2019; Lyammouri & Eddazi, 2020). These campaigns have either been state-sponsored or operated by private, foreign entities. Research from the Africa Center for Strategic Studies has analysed hundreds of misinformation campaigns throughout African countries, showing that 60% of disinformation campaigns are foreign-sponsored. The majority of these state-sponsored campaigns come from Russia, China, and the Gulf States, such as the United Arab Emirates, Saudi Arabia, and Qatar, and are primarily focused on supporting authoritarian governments and military juntas (Africa Center for Strategic Studies, 2024). These campaigns also disproportionately target West African countries such as Burkina Faso, Mali, and Nigeria. West Africa accounts for 43% of targeted disinformation campaigns, followed by East Africa (20%), Southern Africa (15%), Central Africa (13%), and North Africa (9%).

## AI case studies

Given the recent cascade of generative AI tools, understanding of the usage and impact of AI within African elections continues to emerge. Reports and surveys have shown that African governments are leveraging AI to assist with voter registration, authentication, management, and engagement, along with detecting cyber threats and aiding in oversight and decision-making (Itodo, 2024). From 2023-2025, an estimated 36 African countries will hold general or presidential elections to appoint a new head of state. Over the elections and recent political events that occurred before May 2024, there have been many suspected and confirmed cases of AI-generated propaganda influencing electoral outcomes or precipitating coups in countries like Nigeria, Burkina Faso, and Senegal, underscoring the need for comprehensive documentation of generative AI use cases in democratic processes throughout Africa. Recent work by Rest of World, a global news platform, has begun to make headway towards tracking AI-generated election content globally. However, this work only covers one African country, South Africa[1]. There is also a strong need for research examining the spread of generative AI-produced disinformation and its respective impacts within and outside of Africa, given that the majority of reports on such usage are relegated to short-form news stories. To help make headway towards this issue, this section presents case studies illustrating the current usage of generative AI for election and political-related propaganda in Africa.

### Multimedia AI-generated election propaganda

The Nigerian 2023 presidential election was likely one of the first elections within the African continent to experience a deluge of AI-generated content. There were multiple cases of fake endorsements from prominent Nigerians, Hollywood actors, celebrities, and current and former United States presidents (Awojulugbe, 2022; Owoeye, 2022; Adeyemi, 2023). There were also two prominent cases of AI-generated audio, one shared hours before the election in February 2023, purportedly depicting a candidate's plan to rig the election (Shibayan, 2023) and another shared over a month after the elections claimed to depict a call between another presidential candidate and a prominent Nigerian religious figure, which is still under contention as being AI-generated. Electoral AI-generated propaganda can also be disseminated even after elections take place, as seen in Nigeria and Senegal. After the election of Senegal's president, Bassirou Diomaye Faye, in March 2024, a video purporting to depict the president giving a speech was widely spread. However, this video depicted the president's mentor, Ousmane Sonko, giving a speech at a press conference in 2021. However, fact-checkers uncovered that generative AI technology was used to translate his original speech, spoken in French, to appear as if he was speaking English (Kabweza, 2024).

South Africa held their 2024 elections on May 29th. Throughout their election period, Rest of World has tracked at least four cases of generative AI use for political campaigning. The first, detected in February 2024, was a TikTok video posted on a South African meme account appearing to show the American rapper Eminem endorsing the third-largest political party in South Africa, the Economic Freedom Fighters. In March 2024, a video depicting United States President Joe Biden was shared on TikTok, Facebook, X, and WhatsApp, claiming that the United States would enact sanctions on South Africa if the African National Congress, the incumbent party, won the election. Also in March, another video posted on TikTok appeared to show Donald Trump endorsing uMkhonto we Sizwe, a paramilitary party. Rest of World has also shown that the South African Referendum Party has frequently used AI-generated imagery in campaign material since November 2023.

---

[1] https://restofworld.org/2024/elections-ai-tracker

### Political AI-generated propaganda influencing coups

Since 2020, 9 military coups have occurred within West, Central, and East African countries (Green, 2023; Vines, 2024). Countries affected by these coups include Mali (2020, 2021), Guinea (2021), Burkina Faso (January and September 2022), Niger (2023), Gabon (2023), Chad (2021), and Sudan (2021). There have also been a number of attempted coups during this period in Niger, Gabon, Sudan, Guinea-Bissau, The Gambia, Sudan, Sierra Leone, Burkina Faso, São Tomé and Príncipe, and the Democratic Republic of Congo (Duzor & Williamson, 2023). The most recent coup took place in May 2024, when security forces foiled a coup attempt by the Congolese opposition party in the Democratic Republic of Congo. Political propaganda has been shown to impact coups before, during, and after they happen. One of the earliest cases of generative AI propaganda in Africa occurred in December 2018, when a video claiming to depict Ali Bongo, Gabon's president at the time, stirred concern on social media due to unnatural expressions and posture (Breland, 2019). Reports claim that this video raised concerns about his ability to rule and sparked a coup in January 2019, which later failed (Patil & Gori, 2023). Burkina Faso's most recent coup in September 2022 was followed by a series of AI-generated videos from American-based Pan-Africanists appearing to express support for the new military regime. These videos were initially detected on Twitter and Facebook and were shared by users who had received these videos on WhatsApp. While an investigation from VICE confirmed that the videos were created using a tool from the company "Synthesia," the company would not confirm who created the videos, stating that the user had been banned from the platform (Smith Galer, 2023). While the BBC reported a large amount of disinformation shared after Niger's coup in July 2023, there is insufficient evidence to understand the role that generative AI may have played in producing this content, given that mostly older content and manipulated images and videos were shared (Mwai, 2023).

## Understanding efforts toward combating disinformation

### Fact-checking organisations in Africa

Over the past decade, independent fact-checking organisations have been established throughout Africa. These organisations often leverage data-driven practices to advance their work, which has had an impact on increasing media transparency across the continent (Cheruiyot & Ferrer-Conill, 2018). Africa Check, the oldest and most prominent fact-checking organisation in Africa, was founded in 2012. Africa Check operates across South Africa, Kenya, Nigeria, and Senegal have worked diligently to fact-check widely circulated memes, doctored social media content, and claims made by political figures. Africa Check[2] screens around 100,000 inquiries from media consumers yearly and has partnered with platforms such as Facebook as an independent fact checker. Africa Check covers local African languages like Igbo, Yoruba, Hausa, Swahili, Wolof, Afrikaans, Zulu, Setswana, Sotho, Northern Sotho and Southern Ndebele. Other organisations, like PesaCheck[3], established in 2016 in Kenya, focus on verifying financial and statistical information by public figures in East Africa, unpacks budget and census data, and builds AI tools to verify claims automatically. In 2018, the Center for Journalism and Innovation Development established Dubawa[4] in Nigeria, which operates in other English-speaking West African countries, including Ghana, Sierra Leone, Liberia, and The Gambia. Within Africa, other organisations include Zim Fact[5] in Zimbabwe, Moz Check in Mozambique, 211 Check[6] in South Sudan, Beam Reports[7] in Sudan, FactSpace West Africa in Ghana[8], The Stage Media[9] in Liberia, Akhbar Meter[10] in Egypt, and Congo Check[11] in the Democratic Republic of Congo. All of these organisations are members of the International Fact-Checking Network, a global initiative by Poynter, a non-profit media institute and research organisation, to advocate for factual information and fight against misinformation.

---
[2] https://africacheck.org/
[3] https://pesacheck.org/about

### Media literacy educational efforts in Africa

Throughout the continent, fact-checking organisations have been essential in advancing media literacy training for journalists and the general public. Organisations like Dubawa run the Kwame Karikari Fact-checking Fellowship. This 6-month program provides technical training on fact-checking and project mentorship to full-time employees of media organisations in anglophone West Africa. This program also aims to help fellowship recipients establish fact-checking capabilities within their respective organisations. In March 2024, UNESCO supported the Media Academy for Journalism and Communication to host fact-checking and media verification training for 25 journalists in The Gambia (UNESCO, 2024). Other organisations such as Africa Check, the Code for Africa African Fact-Checking Alliance, Inform Africa, and the Media Foundation for West Africa frequently host workshops for media organisations across the continent. The vast majority of fact-checking training programs in Africa have focused on researchers, journalists, or employees of media organisations. This presents a gap in how such efforts reach the general public. While fact-checking training can help journalists conduct their work with the highest integrity, it will be important to develop programs that can effectively target the general public and focus on audiences with lower levels of traditional literacy, those in rural areas, and older adults.

### Tech companies' efforts against disinformation

Along with employing independent fact-checkers to help support efforts to handle disinformation, social media companies have enacted several measures to identify, decelerate, and eradicate such content over the years. Prompted by the increase in message forwarding during the COVID-19 pandemic, much of which was disinformation, Meta introduced global limits on message forwarding in April 2020. Meta reported that these measures helped decrease the spread of highly forwarded messages by 70% (Singh, 2020). WhatsApp had already enacted prior limits on message forwarding due to a series of murders and lynchings in India that occurred from widely spread rumours on the platform (WhatsApp, 2018; Arun, 2019). On their social media platforms, Facebook, Instagram, and Threads, Meta reduces the distribution of content marked by independent fact-checkers as "False", "Altered", or "Partly False" (Meta Business Help Center, 2024), and X (formerly known as Twitter) has enabled features such as Community Notes to allow collaborative labelling of misleading posts (X Corp, 2023). X also reduces incentives for paid subscribers who post tweets fact-checked with Community Notes and has enabled measures to automatically add notes to future images that match previously tagged misleading images (X Corp. (2023). However, this tool is prone to manipulation by groups of contributors who can alter the visibility of tagged content (X Community Notes, 2024; Elliott & Gilbert, 2023). While platforms like TikTok employ automated techniques and partner with over a dozen accredited fact-checking partners that review content in more than 30 languages (Keenan, 2022), these methods may be less effective due to the format of content on their platform, which often makes it more challenging to identify misleading information (Lundy, 2023). While companies and researchers have also explored watermarking techniques to identify AI-generated content (Hwang, 2023), many of these these efforts aren't effective, highlighting a need to develop multifaceted mitigation approaches (Solaiman, 2024).

---

[4] https://dubawa.org/about-us/
[5] https://zimfact.org/
[6] https://211check.org/
[7] https://www.beamreports.com/
[8] https://ghanafact.com/
[9] https://tsmliberia.com/
[10] https://akhbarmeter.org/
[11] https://congocheck.net/

### Leveraging AI tools to combat disinformation

AI-assisted measures to identify disinformation and slow its spread have grown in prominence (Santos, 2023). While many of these methods are developed through academic research, social media companies have increasingly leveraged AI tools to automatically detect disinformation in tandem with more manual processes, such as employing manual fact-checkers. While AI could help scale automated fact-checking, several issues exist with these methods. The majority of LLMs represent a small portion of languages spoken worldwide. If employed as a primary measure to identify disinformation, this may result in inaccurate translations that could falsely identify harmless content as disinformation and inadvertently let harmful content spread. AI datasets and models have also been shown to underrepresent non-Western cultures (Shankar et al., 2017; De Vries et al., 2019; Prabhakaran et al., 2022; Nwatu et al., 2023; Ghosh & Caliskan, 2023), and automated image-verification tools may not have been trained on sufficiently representative datasets to understand cultural nuances. This could exacerbate existing issues with automated content moderation, where users have falsely had their social media accounts deleted due to deficiencies in language translation and misunderstandings of cultural contexts (Shahid & Vashistha, 2023). These issues may also impact how audio-based disinformation is spread, given that automated content moderation systems have a limited understanding of low-resource languages (Nicholas & Bhatia, 2023).

## Navigating the impact of generative on democratic processes in Africa

Generative AI has a strong potential to impact democratic processes, primarily in elections. Within African countries, elections are often plagued by high rates of disinformation, social unrest (Salehyan & Linebarger, 2015; Goldsmith, 2015), internet blackouts (Mpofu, 2022), and ballot tampering (Friese, 2019; Mapuva, 2013). Many of these factors impact civic participation and undermine democracy. The increasing ubiquity of generative AI heightens concerns that AI-fueled disinformation campaigns could

Okolo | 8

exacerbate current issues affecting democratic processes within the African continent. While measures such as internet blackouts could have the unintended positive effect of slowing down the spread of AI-generated disinformation, this content could continue to proliferate in other ways. As countries worldwide continue to grapple with the new complexities posed by AI, especially regarding its impact on democracy, there is a need for coordinated efforts between academia, civil society, governments, and industry to develop novel sociotechnical methods to address these issues. Combating the spread of misinformation will take coordinated efforts from government, civil society, and academic stakeholders to research pressing issues around disinformation, develop novel methods to identify and slow disinformation, help educate and increase public media and AI literacy, and enforce regulatory measures on the production and dissemination of disinformation.

### Issues with disinformation spread

Disinformation is often spread through social media and digital platforms and can be hard to decelerate, given the ability of these platforms to amplify information at scale. The prior section has discussed a number of strategies employed by digital platforms to hinder disinformation. However, issues with the robustness of platform policies and the effectiveness of automated and manual methods persist. While companies such as Meta have enacted a variety of safeguards to identify disinformation and limit its spread, users intent on spreading information can thwart these measures relatively easily (e.g., copying and pasting a WhatsApp message to circumvent forwarding limits). Measures to stop disinformation from spreading through platforms and services that leverage end-to-end encryption, such as WhatsApp, Messenger, Telegram, and Signal, may also be complex to implement due to the respective technical requirements around encryption. Independent fact-checking has also arisen as a popular measure to decelerate the spread of disinformation. However, given that fact-checking is often a very detailed and manual process, efforts to slow down the spread of disinformation and present verified information may not reach impacted consumers in a timely manner. This can have extreme implications regarding disinformation in sensitive contexts, such as dangerous health remedies, false accusations, and social conflict. The offline transference of misinformation can exacerbate these issues as it shifts from digital platforms to print media to word of mouth. An emerging area of research is focused on examining mis/disinformation spread in communities with low internet penetration (Gadjanova et al., 2022), which will become increasingly important to understand as researchers and companies broaden their mitigation methodologies.

### Improving AI, digital, and media literacy through citizen engagement

Research has shown that citizens with limited digital and media literacy often inadvertently disseminate propaganda (Rubin, 2019). However, there are instances, such as the 2019 Nigerian elections, where citizens intentionally shared disinformation about specific presidential candidates, knowing this information was false (Hitchen et al., 2019). Research has also shown that AI-generated content is often difficult to detect (Sadasivan et al., 2023; Lu et al., 2023). While the research is mixed on the specific demographics of populations who are more likely to spread mis/disinformation, older adults and those with lower educational attainment are less likely to identify disinformation (Full Fact et al., 2020; Unfried & Priebe, 2024). In Nigeria specifically, reports have indicated that younger adults often received WhatsApp broadcasts about viral protection and remedies from older relatives during the 2014 Ebola outbreak (Kazeem, 2019). These nuances, combined with the increasing digital divide in Africa, necessitate the development of efforts that ensure marginalised populations such as those in rural areas, older adults, and women are equally involved in educational activities against disinformation.

To help mitigate many of the impacts of disinformation, organisations, and governments should focus on increasing AI, digital, and media literacy of the general public. Given the concentration of fact-checking initiatives dedicated to serving media stakeholders, such efforts could be adapted and expanded to meet the needs of communities across the continent. Civil society and non-profit organisations will also have a prominent role in supplementing existing efforts to increase media and AI literacy. To help support the increasing number of efforts focused on curbing the spread of AI-generated disinformation, governments must increase funding for these initiatives. While there may be limited funds for external expenditures, governments can also seek partnerships with companies, multinational agencies, and foundations to support these initiatives. Investing in such efforts would also scale the significant work being conducted by fact-checking organizations across the continent. While issues of AI, digital, and media literacy are complex, sufficient funding of educational programs, systematised approaches to collaboration, and dedicated research agendas to examine best strategies for upskilling can help make a significant difference.

### Increasing governmental and regulatory efforts toward combating disinformation

Since bad actors versus specific individuals within the general public are often the primary architects of disinformation campaigns, dedicated efforts are needed to control the spread of disinformation through these respective means. Engaging citizens in AI, digital, and media literacy efforts will only be effective if measures are taken to stop disinformation at its source. The ability of bad actors to evade detection and mitigation techniques set in place by digital platforms also heightens the importance of improved technical standards to increase the efficacy of these techniques and increased regulation to ensure compliance and accountability for violations. At the same time, the majority of countries within Africa have passed data privacy and protection measures. With an estimated 36 out of 54 African countries having enacted comprehensive data protections, comprehensive continental-wide frameworks are needed to address the unique data issues posed by generative AI. While the African Union (AU) Convention on Cyber Security and Personal Data Protection (also known as the Malabo Convention) aims to establish comprehensive norms and regulations for cybersecurity and data privacy across the African continent, only 15 countries have ratified this treaty, posing concerns about the ability of the AU to enforce the outlined protections. African governments will also have to establish regulatory frameworks that can effectively govern the use of generative AI and other emerging technologies while promoting safeguards to prevent the spread of disinformation.

## Conclusion

While the landscape of AI-generated disinformation is still evolving within Africa, this emerging technology poses severe threats to African democracy. The preliminary impact of generative AI on elections in Africa shows serious indications of what will likely come for subsequent elections within the continent as this technology evolves over the next few years. With additional concerns about the impact of disinformation on gender equality, conflict prevention, and national security, African governments must address these issues and support civil society, academic, and industry stakeholders to develop robust solutions. This paper reviews the state of AI-generated disinformation within African countries to understand how African governments might effectively counteract the impact of AI-generated propaganda and propose tangible recommendations stakeholders can adopt to address these multifaceted challenges.